\documentclass[rnote]{aa} % for the research notes
\usepackage{graphicx,natbib}
\usepackage{txfonts}
\newcommand{\Msun}{\ensuremath{\mathit{M}_{\odot}}}
\newcommand{\sh}{\mathrm{sh}}

\newcommand{\ej}{\mathrm{ej}}
\newcommand{\csm}{\mathrm{csm}}

\begin{document} 

   \authorrunning{Moriya}
   \titlerunning{`snow-plow' phase}

   \title{
On the `snow-plow' phase of supernovae interacting with dense
circumstellar media
 }

   \subtitle{}

   \author{Takashi J. Moriya
          \inst{1,2}
          }

   \institute{
Argelander Institute for Astronomy, University of Bonn,
Auf dem H\"ugel 71, D-53121 Bonn, Germany
   \\
              \email{moriyatk@astro.uni-bonn.de}\and
Research Center for the Early Universe, Graduate School of Science,
University of Tokyo, Hongo 7-3-1, Bunkyo, Tokyo, Japan
             }

%   \date{Received September 15, 1996; accepted March 16, 1997}
   \date{}

\abstract{
Recently, \citet{moriya2013} developed an analytic bolometric light
curve model for supernovae interacting with dense circumstellar media.
Because of the dense circumstellar medium, the shocked region is
assumed to be radiative and make a thin dense shell.
The model is based on the conservation of momentum in the shocked dense shell.
However, the analytic model was mentioned to neglect 
the `snow-plow' phase of the shocked dense shell by \citet{ofek2014}.
The `snow-plow' or momentum-conserving phase refers to the period in which
the momentum injection from
the supernova ejecta is almost terminated and the radiative shocked
dense shell keeps moving due only to the momentum previously provided
by the supernova ejecta.
In this Note, I clarify that the analytic model of \citet{moriya2013} does take
the `snow-plow' phase into account
and the criticism of \citet{ofek2014} is incorrect.
In addition, \citet{ofek2014} related the sudden luminosity break
observed in the light curve of Type IIn SN 2010jl to the transition to the
`snow-plow' phase. However, I argue that the sudden transition to the
`snow-plow' phase is not consistent with the luminosity break
observed in SN 2010jl.
The luminosity break is likely to be related to other phenomena
like the dense shell exiting the dense part of the circumstellar medium.
}

% \abstract{}{}{}{}{} 
% 5 {} token are mandatory
 
%  \abstract
%  % context heading (optional)
%  % {} leave it empty if necessary  
%   {context}
%  % aims heading (mandatory)
%   {aims}
%  % methods heading (mandatory)
%   {methods}
%  % results heading (mandatory)
%   {results}
%  % conclusions heading (optional), leave it empty if necessary 
%   {conclusions}


   \maketitle
%
%________________________________________________________________

\section{Introduction}\label{introduction}
Mass loss of massive stars has critical roles in many aspects of
astrophysical phenomena \citep[e.g.,][]{smith2014}.
One of the outstanding problems regarding the stellar mass loss
is the existence of supernova (SN) progenitors having
extremely high mass-loss rates immediately before their explosion.
Especially, Type IIn SNe are known to show the signatures of
the dense circumstellar media (CSM) which are presumed to be related
to the extreme mass-loss activities of the progenitors immediately before
their explosion \citep[e.g.,][]{fransson2002,taddia2013,kiewe2012}.
Recently, \citet[][M13 hereafter]{moriya2013} developed a bolometric light
curve (LC) model for SNe interacting with the dense CSM. The model is
used to interpret many Type IIn SN LCs to estimate the general properties of
the mass loss of Type IIn SN progenitors \citep{moriya2014}.
Similar LC models are suggested previously for the CSM from the steady
mass loss \citep[e.g.,][]{svirski2012,wood-vasey2004}
but M13 considered more general cases.
\citet{ofek2014} have also developed a similar LC model to the M13 model
to interpret the LC of SN 2010jl, which is one of the most well-observed
Type IIn SNe so far (see \citealt{fransson2013,ofek2014} and the
references therein).
One interesting feature in the LC of SN 2010jl is the luminosity break
observed at around 320 days after the first detection of the SN \citep{fransson2013,ofek2014}.
\citet{ofek2014} related the break to the transition of the forward
shock propagating in the dense CSM to the `snow-plow' or
momentum-conserving phase \citep{svirski2012}.
If the shocked region by the SN explosion is radiative as is the case in
Type IIn SNe, the shocked region makes a thin
dense shell \citep[e.g.,][]{chevalier1994}.
The `snow-plow' phase corresponds to the phase
when the momentum supply from the SN ejecta to the shell is almost
terminated. Roughly speaking, the shell enters the `snow-plow' phase
when the mass of the shocked dense CSM is
comparable to the mass of the SN ejecta.
\citet{ofek2014} mentioned that the `snow-plow' phase
is neglected in the model of M13.

In this Note, I will first clarify that the criticism of \citet{ofek2014}
on the M13 LC model regarding the `snow-plow' or momentum-conserving
phase is incorrect. I will show that the `snow-plow' phase is properly
taken into account in the M13 model. I confirm this by
following the evolution of the shocked dense shell numerically.
Finally, I discuss the luminosity break observed in SN 2010jl
and argue that the sudden luminosity break observed in SN 2010jl
is inconsistent with the transition to the `snow-plow' phase.

\section{Light curve models for interacting SNe}

\subsection{M13 analytic model}
I briefly summarize the M13 analytic LC model.
More detailed and complete discussion of the model is in M13.
The model assumes that the shocked SN ejecta and shocked dense CSM create
a thin dense shell because of the efficient radiative cooling.
Thus, the model assumes that the location of the shocked region can
be expressed with a single radius $r_\sh(t)$. Under this assumption,
the evolution of $r_\sh(t)$ is governed by
the conservation of momentum, 
\begin{equation}
M_\sh\frac{dv_\sh}{dt}=4\pi r_\sh^2\left[\rho_\ej (v_\ej-v_\sh)^2-\rho_\csm (v_\sh-v_\csm)^2 \right],\label{momentum}
\end{equation}
where $M_\sh$ is the mass of the shell which is the sum of the mass of
the shocked SN ejecta and shocked dense CSM,
$v_\sh$ is the velocity of the shell, 
$\rho_\ej$ is the density of the SN ejecta entering the shell,
$\rho_\csm$ is the density of the CSM entering the shell, and
$v_\csm$ is the CSM velocity. 
The first term on the right-hand side of Equation (\ref{momentum}),
i.e. $4\pi r_\sh^2 \rho_\ej (v_\ej-v_\sh)^2$, represents the momentum provided
by the SN ejecta to the shell. The second term
$-4\pi r_\sh^2 \rho_\csm (v_\sh-v_\csm)^2$
is the momentum provided by the dense CSM.
The M13 LC model is obtained solely by solving Equation (\ref{momentum}).

The SN ejecta is assumed to have two density structures,
$\rho_\ej\propto r^{-n}$ outside and
$\rho_\ej\propto r^{-\delta}$ inside.
The outer part of SN ejecta continues to enter the shell until the time
$t_t$ (see M13 for details). After $t_t$,
the inner part of the SN ejecta starts to enter the shell.
The evolution of the shell before $t_t$
can be followed analytically but the general analytic solution does not
exist after $t_t$.
As is discussed in M13, long after $t_t$ when most of the SN
ejecta has entered the shell, little momentum is provided from
the SN ejecta and the equation for the conservation of momentum can be
approximated as (Equation 10 in M13)
\begin{equation}
M_\sh\frac{d^2r_\sh}{dt^2}=4\pi r_\sh^2 \left[-\rho_\csm
 (v_\sh-v_\csm)^2\right].\label{snowplow}
\end{equation}
Equation (\ref{snowplow}) clearly corresponds to the phase when
the radiatively cooling shell continues to move with the momentum
initially provided
by the SN ejecta without any additional momentum supply from the SN ejecta,
i.e., the `snow-plow' phase.
M13 suggested that the evolution of the shell can be approximated by
Equation (\ref{snowplow}) long after $t_t$ when most of the SN ejecta is
shocked by the shell.

For the case of the steady mass loss ($\rho_\csm\propto r^{-2}$),
Equation (\ref{snowplow})
is shown to have the analytic solution in M13 and they
provided the asymptotic analytic bolometric LC of the interacting SNe
corresponding to this `snow-plow' phase (Equation 29 in M13)
\begin{equation}
L=\frac{\epsilon}{2}\frac{\dot{M}}{v_\csm}
\left(\frac{2E_\ej}{M_\ej}\right)^{\frac{3}{2}}
\left[1+2\frac{\dot{M}}{v_\csm}
\left(\frac{2E_\ej}{M_\ej^3}\right)^{\frac{1}{2}}t\right]^{-\frac{3}{2}},
\label{s=2longafter}
\end{equation}
where $\epsilon$ is the conversion efficiency from the kinetic energy to
radiation, $\dot{M}$ is the mass-loss rate,
$E_\ej$ is the kinetic energy of the SN ejecta, and
$M_\ej$ is the mass of the SN ejecta.
\citet{svirski2012} earlier showed that the bolometric luminosities
of the interacting SNe eventually follow $L\propto t^{-1.5}$
in the `snow-plow' phase.
As can be clearly seen from
Equation (\ref{s=2longafter}), the M13 analytic solution approaches
$L\propto t^{-1.5}$ after the following condition is satisfied
\begin{equation}
\frac{\dot{M}}{v_\csm}
\left(\frac{2E_\ej}{M_\ej^3}\right)^{\frac{1}{2}}t \sim 1. \label{cond1}
\end{equation}
If the radius of the shell can be roughly approximated as
$r_\sh \sim (2E_\ej/M_\ej)^{0.5}t$ as is assumed in \citet{svirski2012},
the condition (\ref{cond1}) corresponds to
\begin{equation}
 M_\mathrm{scsm} \sim M_\ej,
\end{equation}
where $M_\mathrm{scsm}=\frac{\dot{M}}{v_\csm}r_\sh $ is the mass of the
shocked CSM. Thus, the M13 model clearly takes the `snow-plow' phase
into account.
Equation (\ref{s=2longafter}) is shown to be a good approximation
for the luminosity evolution at the transitional phase to the
`snow-plow' phase as well in the next section.

%The mass contained in the outer part of the SN ejecta whose density
%structure follows $\rho_\ej\propto r^{-n}$ is

%The CSM density structure is also assumed to follow power-law
%($\rho_\csm \propto r^{-s}$).

\subsection{Numerical confirmation}\label{numerical}
To confirm that the M13 analytic model properly takes the
`snow-plow' phase into account,
I numerically solve the equation for the
conservation of momentum (Equation \ref{momentum}).
The CSM from the steady mass loss 
with $\dot{M}=0.1$ $M_\odot~\mathrm{yr^{-1}}$ and 
$v_\csm=100$ $\mathrm{km~s^{-1}}$ is put outside the
homologously expanding SN ejecta with the two density structure
components with $n=10$ and $\delta=0$. The SN ejecta mass and energy
are assumed to be 10 \Msun\ and $3\times 10^{51}$ erg, respectively.
The kinetic energy of the SN ejecta is chosen to match the early
bolometric luminosity of SN 2010jl with
$\epsilon=0.5$ (see Section \ref{10jl} for the discussion of SN 2010jl). 

\begin{figure}
\centering
\includegraphics[width=\columnwidth]{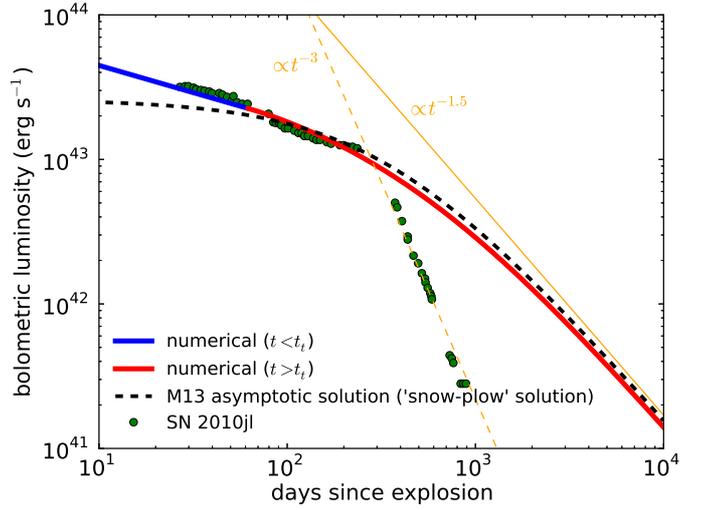}
\caption{
Bolometric LCs of the numerical model and M13 asymptotic model.
The asymptotic model corresponds to the 
`snow-plow' phase (Equation
 \ref{snowplow}).
When the mass of the CSM swept by the shocked shell is sufficiently
larger than the SN ejecta mass,
the two LCs follows $L\propto t^{-1.5}$ as is
 expected in the `snow-plow' phase (\citealt{svirski2012}).
The bolometric LC of SN IIn 2010jl obtained by \citet{fransson2013}
is shown for comparison.
}
\label{LC}
\end{figure}

\begin{figure}
\centering
\includegraphics[width=\columnwidth]{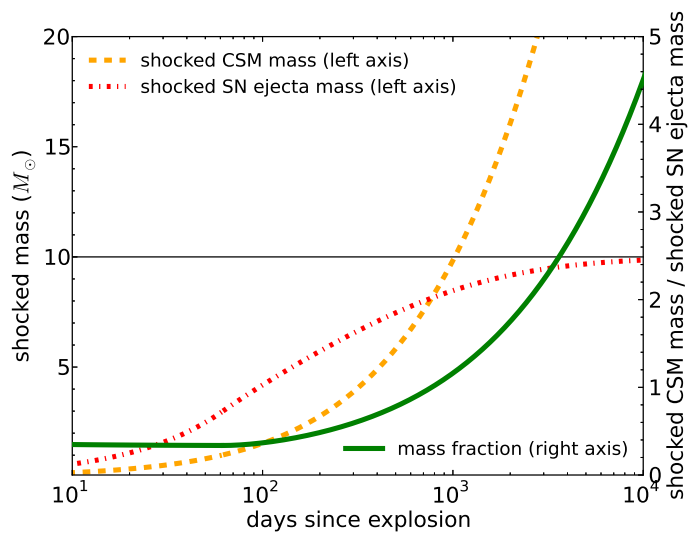}
\caption{
Mass of the shocked CSM and SN ejecta and their fraction obtained by the
 numerical model presented in Section \ref{numerical} and Figure \ref{LC}.
}
\label{fraction}
\end{figure}

Figure \ref{LC} presents the bolometric LC obtained numerically as well
as the asymptotic solution or the solution at the 
`snow-plow' phase derived by M13 (Equation \ref{s=2longafter}).
The blue part of the LC corresponds to the phase when the SN density
structure entering the shocked shell is $\rho_\ej\propto r^{-n}$ $(t<t_t)$
and the red part to $\rho_\ej\propto r^{-\delta}$ $(t>t_t)$.
The transition time $t_t=61$ days expected from the M13 model matches
that obtained in the numerical model. The asymptotic LC model of M13
which corresponds to the `snow-plow' phase (Equation \ref{s=2longafter}) starts to match the
numerical LC relatively soon after $t_t$.
Both analytic and asymptotic models approach $L\propto t^{-1.5}$ in
the `snow-plow' phase as is expected
(\citealt{svirski2012}).
An important feature to note is that there is no sudden transition to
the `snow-plow' phase. The bolometric LC gradually starts to follow
$L\propto t^{-1.5}$ with time as the fraction of the shocked CSM mass to the
shocked SN ejecta mass increases (see Figure \ref{fraction}).

\section{SN 2010jl}\label{10jl}
I have clarified that the M13 bolometric LC model takes the
`snow-plow' or momentum-conserving phase into account in the previous section.
\citet{ofek2014} suggested that the luminosity break observed in
SN 2010jl is related to the transition to the `snow-plow' phase
and the time of the break corresponds to when the mass of the shocked
CSM and SN ejecta gets comparable. However, as we have shown in the
previous section, the transition to the `snow-plow' phase
is not expected to occur with the small timescale observed in SN 2010jl.

\begin{figure}
\centering
\includegraphics[width=\columnwidth]{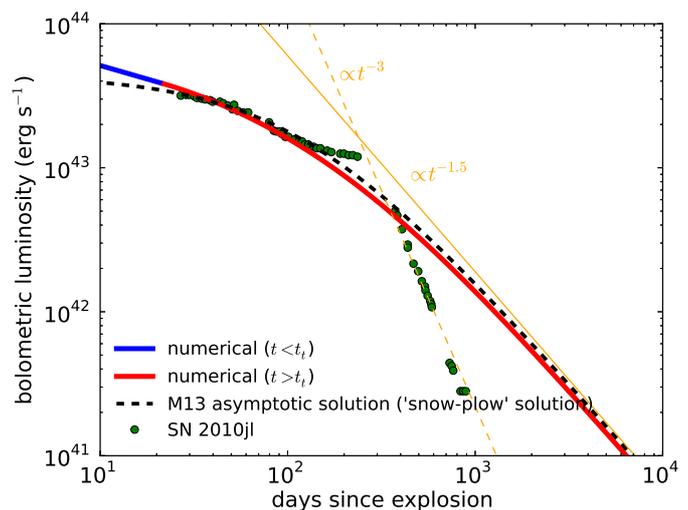}
\includegraphics[width=\columnwidth]{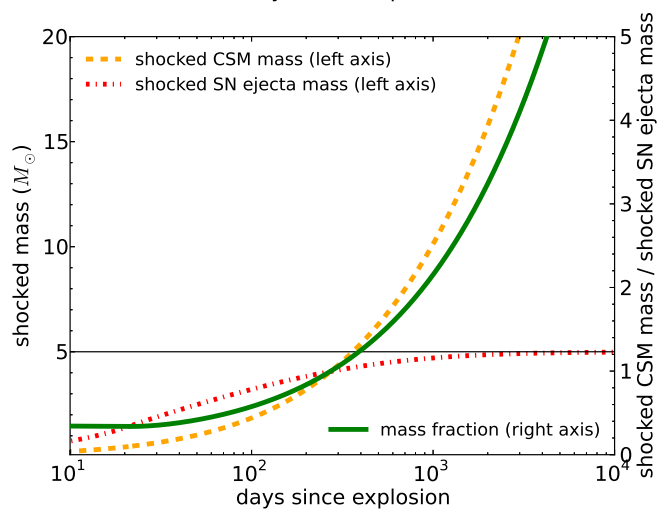}
\caption{
Top: Numerical bolometric LC model in which $M_\ej\sim M_\mathrm{scsm}$
is satisfied
at around 350 days since the explosion, when the luminosity break is
observed in SN 2010jl. Bottom: Evolution of the shocked mass and the
 mass fraction in the model presented in the top panel.
}
\label{another}
\end{figure}

Since the numerical model presented in the previous section satisfies
$M_\ej = M_\mathrm{scsm}$ at around 1000 days since the explosion,
I show another model in which $M_\ej = M_\mathrm{scsm}$ is satisfied
at around 350 days when the luminosity break of SN 2010jl is observed.
Figure \ref{another} shows the result. The ejecta mass is reduced
to $M_\ej=5\ M_\odot$ in the new model to satisfy $M_\ej =
M_\mathrm{scsm}$ at around 350 days.
The conversion efficiency
$\epsilon$ is also reduced to $0.3$ to match the luminosity of SN
2010jl but the other parameters are kept the same.
The transition to the `snow-plow' phase does
not make a luminosity break as is observed in SN 2010jl.

The sudden transition to the `snow-plow' phase may occur if
a high-density circumstellar shell exists on top of the dense CSM considered in
the model. If the shocked shell suddenly encounters a high-density (and thus
massive) shell, the shocked shell can be suddenly decelerated and
can suddenly turn to the `snow-plow' phase.
However, when the shocked shell encounters the high-density shell,
a sudden bolometric luminosity increase is expected
rather than a break in the luminosity. This is because the increased density
makes the deceleration of the shocked shell more efficient and
more kinetic energy will be converted to radiation.
Thus, the sudden transition to the `snow-plow'
phase by the collision to the high-density shell
should be accompanied by the sudden luminosity increase which is
not observed in SN 2010jl. This means that the sudden luminosity break
observed in SN 2010jl is not likely to be related to the transition to the
`snow-plow' phase.

Alternative causes for the sudden luminosity decline in SN 2010jl were
also discussed in \citet{ofek2014,fransson2013}. One possibility is 
that the shocked shell gets out of the dense part of CSM at the time of
the transition. Also, the CSM density slope may have suddenly become steep.
As is noted by the previous works, the luminosity follows $L\propto t^{-3}$
after the break but no clear reason for this has been suggested.
The CSM density structure should be steeper than
$\rho_\csm \propto r^{-3}$ if the break is due to the transition to the
different density slope (M13).

\section{Conclusion}
I have shown that the analytic LC model of M13 which is developed to
model SNe interacting with dense CSM takes the 'snow-plow' phase, or
momentum-conserving phase, of the shocked dense shell into account.
The model was criticized for neglecting it by \citet{ofek2014}
but the criticism is incorrect.
The M13 LC model can reproduce the LC behavior expected in
the `snow-plow' phase well ($L\propto t^{-1.5}$ in the steady
wind, see \citealt{svirski2012}).
The analytic LC model is also compared to the numerical model and
the analytic model is shown to reproduce the numerical result well.

The luminosity break observed in SN 2010jl was related to the transition
to the `snow-plow' phase by \citet{ofek2014}. However,
the sudden luminosity break observed in SN 2010jl is not expected from
the transition
to the `snow-plow' phase. The transition is expected to occur
gradually as the shocked shell accumulates the CSM mass.
The sudden transition may occur if there is a high-density shell in
the CSM but the collision of the shocked shell to the high-density shell
is expected to be accompanied by the luminosity increase, not the break.
Thus,
the luminosity break is not likely to be related to the transition to
the `snow-plow' phase.

\begin{acknowledgements}
TJM is supported by the Japan Society for the Promotion of Science
Research Fellowship for Young Scientists (23\textperiodcentered 5929).
\end{acknowledgements}

\end{document}